# Energy Efficient Water-Filling Power Allocation with Cell Division for Massive MIMO Systems

Abdolrasoul Sakhaei Gharagezlou
Department of Electrical & Computer Engineering
University of Tabriz
Tabriz, Iran
abdolrasoulsakhaei@gmail.com

Jafar Pourrostam
Department of Electrical & Computer Engineering
University of Tabriz
Tabriz, Iran
j.pourrostam@tabrizu.ac.ir

Mahdi Nangir
Department of Electrical & Computer Engineering
University of Tabriz
Tabriz, Iran
nangir@tabrizu.ac.ir

Mir Mahdi Safari
Department of Electrical & Computer Engineering
University of Tabriz
Tabriz, Iran
safari@tabrizu.ac.ir

*Abstract*— In this paper, we consider the power allocation problem for the downlink of the massive multiple-input multiple-output (MIMO) systems. We propose a new scheme by exploiting the water-filling algorithm in a cell with two zones. It is allocated more power to the farther users, and also users with better channel conditions receive more power in each zone. The users with better channel gain have a higher priority than others in the energy efficiency (EE) point of view. Combining the water-filling and cell division techniques in the MIMO systems leads to a reach EE maximization. We also use standard interference function (SIF) to propose a new iterative algorithm for solving the optimization problem and obtain an efficient power allocation scheme. Simulation results show that the proposed algorithm outperforms other similar algorithms from the EE point of view.

## I. Introduction

With the advent of portable and accessible smart devices, the next generation of cellular systems meets a significant increase in connecting devices to the internet. Through the past ten years, wireless communication has experienced a transfer in the multiple-input multiple-output (MIMO) communication systems from a theoretical concept to a practical technique which increases the performance of wireless networks. The use of MIMO in telecommunication systems leads to four considerable performance improvements, including arrays increasing, diversity increasing, spatial multiplexing increasing, and interference reducing. Reliable communication, high energy efficiency (EE), and low-complexity performance in the signal processing are other advantages of the MIMO systems. In general, the MIMO systems are divided to two categories, which include the point-to-point or single-user MIMO systems (SU-MIMO), and the multiple-user MIMO systems (MU-MIMO). As the number of antennas at the base station (BS) increases, the destructive effects of the channel and intercellular interference are eliminated. When the number of antennas in the BS is doubled, the transmission power is reduced to half of the original power in this system. This factor doubles EE [1]. If the number of BS antennas increases ($\geq 8$), then the system is called a massive MIMO. Massive MIMO is one of the underlying technologies for the new 5G and beyond, which has been explored in recent years [2]. In the massive MIMO cellular communication systems, the BS is equipped with a large number of antennas and simultaneously serves multiple users [2,3]. It is worth mentioning that the power allocation (PA) problem has a vital role in determining and exploiting the capacity of MIMO and massive MIMO systems because of the existence of numerous antennas at the transmitter and receiver side, while in the single-input single-output (SISO) case, the PA is not an important issue. Recently some research works are accomplished to solve the PA problem in the massive MIMO systems. In [4], the authors have solved the PA problem under the constraint of minimum required power for each user in the massive MIMO system. The authors have obtained a closed form solution for the PA problem using Karush-Kuhn-Tucker (KKT) method and their results show performance improvement. An optimal power allocation scheme for maximizing system throughput using the successive interference cancellation (SIC) method for the non-orthogonal multiple access (NOMA) system is presented in the downlink [5]. In the downlink of the NOMA systems, signals of each user are mounted and mixed on the transmitter side and the SIC is implemented on the receiver side to separate them. It is shown that the NOMA method performs better than the orthogonal multiple access (OMA) method from the system throughput point of view.

In [6], authors have proposed the energy-efficient PA for an MU-massive MIMO system by using the standard interference function (SIF) method in a single-cell scenario. In the article [7], the authors present an EE allocation scheme for a massive MIMO-NOMA system. In that scheme, the cell is divided into two zones. The first area belongs to users whose distance to the base station is less than half the radius of the cell, and the second area includes users whose distance to the base station is more than half the radius of the cell. Also according to the total distance of users in that area, the maximum transmission power for each area is determined. In paper [8], the energy-efficient PA for a MIMO-NOMA system with multiple users in a cluster is investigated. To ensure the quality of user service, the minimum rate required by users is pre-defined. Due to the quality of service (QoS) required by users, first, it is necessary to compare the total transfer power with the total power required by all users to determine whether the intended optimization problem is feasible or not. If possible, a closed-form solution is provided for the problem of maximizing the total data rate of users, and accordingly, the problem of maximizing EE is solved according to non-convex fraction planning. Otherwise, a low-complexity user acceptance scheme is presented that sorts users one by one in ascending order of the power required to meet the QoS.

On the other hand, to design efficient communication systems, the capacity-achieving water-filling scheme has been considered in many works [9-12]. In [13], a low-complexity iterative water-filling (LC-IWF) in an LTE-advanced network MIMO (Net-MIMO) is proposed to achieve further enhancement. The LC-IWF algorithm is exploited to allocate bandwidth optimally and power among the participating users in the network. In this paper, we propose an energy-efficient water-filling PA scheme in the massive MIMO-NOMA system. By dividing each cell into two zones, we can allocate more power to the farther users, and by doing the water-filling algorithm we can allocate more power to the users in each zone that has better channel gain and condition. This consideration is important because users with better channel gains have a higher priority than their counterparts.

The rest of the paper is organized as follows. The system model is introduced in Section II. The proposed water-filling based PA strategy is elaborated in Section III. Simulation results are shown in Section IV, and conclusions are finally drawn in Section V.

## II. SYSTEM MODEL

We consider the downlink MU-Massive MIMO system in which the BS has $M$ antennas and $Z$ users in each cluster with $N$ antennas. The total number of users in the system is $M \times Z$, which are grouped into $M$ clusters randomly with $Z$ users per cluster. For user $(m,z)$, $m \in \{1,...,M\}$, $z \in \{1,...,Z\}$, the channel matrix between the BS and the $z$-th user in the $m$-th cluster, i.e., $H_{m,z} \in \mathbb{C}^{N \times M}$ is assumed to be quasi-static independent and identically distributed (i.i.d.). For the considered MU-Massive MIMO system, the corresponding transmitted signals from the BS can be derived as:

$$\mathbf{s} = \mathbf{P} \begin{bmatrix} \sqrt{P_{\max} \rho_{1,1}} x_{1,1} + ... + \sqrt{P_{\max} \rho_{1,Z}} x_{1,Z} \\ \sqrt{P_{\max} \rho_{M,1}} x_{M,1} + ... + \sqrt{P_{\max} \rho_{M,Z}} x_{M,Z} \end{bmatrix}, \quad (1)$$

where $\mathbf{P} \in \mathbb{C}^{M \times M}$ the BS pre-coding matrix and $P_{\max}$ is the total transmission power for the BS. Signal and power allocation coefficient for the user $(m,z)$ are shown as $x_{m,z}$ and $\rho_{m,z}$ respectively. It is worth mentioning that the power allocation coefficients must meet $\sum_{m=1}^{M} \sum_{z=1}^{Z} \rho_{m,z} \leq 1$.

The received signal at the user $(m, z)$ is expressed as:

$$\mathbf{y}_{m,z} = \mathbf{H}_{m,z} \mathbf{P} \begin{bmatrix} \sqrt{P_{\max} \rho_{1,1}} x_{1,1} + ... + \sqrt{P_{\max} \rho_{1,Z}} x_{1,Z} \\ \sqrt{P_{\max} \rho_{M,1}} x_{1,1} + ... + \sqrt{P_{\max} \rho_{M,Z}} x_{M,Z} \end{bmatrix} + \mathbf{n}_{m,z}, \quad (2)$$

where $\mathbf{n}_{m,z}$ denotes the noise parameter $CN(0, \sigma^2 I)$. By applying the detection vector $\mathbf{vd}_{m,z}$ to the observed signal, from (2), we have [8]:

$$\mathbf{vd}_{m,z}^H \mathbf{y}_{m,z} = \mathbf{vd}_{m,z}^H \mathbf{H}_{m,z} \mathbf{p}_m \sum_{z=1}^{Z} \sqrt{P_{\max} \rho_{m,z}} x_{m,z} \quad (3)$$
$$+ \sum_{i=1, i \neq m}^{M} \mathbf{vd}_{m,z}^H \mathbf{H}_{m,z} \mathbf{p}_i \mathbf{x}_i + \mathbf{vd}_{m,z}^H \mathbf{n}_{m,z},$$

where $\mathbf{x}_k$ denotes the $k$-th row of $\mathbf{x}$ in $\mathbf{x}=\mathbf{Ps}$. The second term in (3) is indicate the interference from other clusters which is omitted because of $\mathbf{vd}_{m,z} \mathbf{H}_{m,z} \mathbf{p}_i = 0$ for any $i \neq m$, based on the assumptions in [8]. By considering the effective channel gain order [14]:

$$\left| \mathbf{vd}_{m,1}^H \mathbf{H}_{m,1} \mathbf{p}_m \right|^2 \geq ... \geq \left| \mathbf{vd}_{m,Z}^H \mathbf{H}_{m,Z} \mathbf{p}_m \right|^2, \quad (4)$$

The achieved data rate at user $(m,z)$ is expressed as:

$$R_{m,z} = \log_2(1 + \frac{\zeta \rho_{m,z} \left| \mathbf{vd}_{m,z}^H \mathbf{H}_{m,z} \mathbf{p}_m \right|^2}{1 + \zeta \sum_{j=1}^{z-1} \rho_{m,j} \left| \mathbf{vd}_{m,z}^H \mathbf{H}_{m,z} \mathbf{p}_m \right|^2}), \quad (5)$$

where $\zeta = \frac{P_{\max}}{\sigma^2}$ denotes the transmission signal-to-noise ratio (SNR). We define the EE of the system as:

$$EE = \frac{\sum_{m=1}^{M}\sum_{z=1}^{Z} R_{m,z}}{P_c + P_{\max}\sum_{m=1}^{M}\sum_{z=1}^{Z} \rho_{m,z}}, \quad (6)$$

where $P_c$ is the fixed circuit power consumption. The term $P_{\max}\sum_{m=1}^{M}\sum_{z=1}^{Z}\rho_{m,z}$ is related to flexible transmission power, so the total power consumption is consist of fixed circuit power consumption and flexible transmission power.

## III. THE PROPOSED POWER ALLOCATION ALGORITHM

In this section, the proposed energy efficient water-filling PA with cell division scheme is formulated and then a method for solving the problem is explained.

### A. Optimization problem formulation

We aim to maximize the EE of the system. The objective function is in a fractional form and is a non-convex problem, hence obtaining an optimal solution is non-trivial. To solve it in a tractable way, we first turn it to the corresponding Spectral Efficiency (SE) maximization problem. According to the definition of EE in (6), to maximize the EE, we need to maximize the corresponding SE under the assumption of dividing the cell into two areas and ignoring the interference between users in a cell. The considered problem can be formulated as:

$$\max_{\{\rho_{m,1},\rho_{m,2},\dots,\rho_{m,z}\}} \sum_{z=1}^{Z} \log_2(1+\rho_{m,z}c_z), \quad (7.a)$$

$$C_1: P_{\max}\sum_{z:d_z<\frac{D}{2}} \rho_{m,z} = \alpha P_T, \quad (7.b)$$

$$C_2: P_{\max}\sum_{z:d_z>\frac{D}{2}} \rho_{m,z} = (1-\alpha)P_T, \quad (7.c)$$

where $P_T$ denotes the flexible transmission power and $c_z$ represents the ratio of the signal to the noise. $\alpha$ denotes the power allocation coefficient for the first zone, and $D$ is the radius of the cell. Constraint (7.b) indicates that for users in the first zone, they are less than half the radius of the cell, $\alpha$ times the total power allocated. The constraint (7.c) has a same similar interpretation.

The Lagrangian function method can be used for solving the (7.a) convex problem. Let $\Phi$ be the Lagrangian function of (7), then it can be formulated as follows:

$$\phi(\mathbf{p},\lambda_1,\lambda_2) = -\sum_{z=1}^{Z}\log_2\left(1+\rho_{m,z}c_z\right) - \lambda_1\left(\alpha P_T - \sum_{z:d_z<\frac{D}{2}} p_{m,z}\right) \\ -\lambda_2\left((1-\alpha)P_T - \sum_{z:d_z>\frac{D}{2}} p_{m,z}\right). \quad (8)$$

$\lambda_1$ and $\lambda_2$ are the Lagrangian multipliers corresponding to the transmission power constraint of the first zone and the second zone, respectively. The necessary and sufficient condition to obtain the optimal transmission power in the first and second zones are respectively formulated as follows,

$$\frac{\partial \phi}{\partial \rho_{m,z}} = \frac{c_z}{1+\rho_{m,z}c_z} - \lambda_1 = 0, \quad \rho_{m,z} = \left(\frac{1}{\lambda_1} - \frac{1}{c_z}\right)^+, \quad (9)$$

$$\frac{\partial \phi}{\partial \rho_{m,z}} = \frac{c_z}{1+\rho_{m,z}c_z} - \lambda_2 = 0, \quad \rho_{m,z} = \left(\frac{1}{\lambda_2} - \frac{1}{c_z}\right)^+.$$

The parameters that are most important in allocating full transmission power to two zones are the number of users in that area and the distance to the base station [3]. Note we have

$$\alpha = \frac{\sum_{z:d_z<\frac{D}{2}} d_z^2}{\sum_{z=1}^{Z} d_z^2}. \quad (10)$$

Here, we provide our proposed algorithm with detail in a step by step procedure. Note that, $\theta_1$ and $\theta_2$ are positive steps which are set heuristically during the algorithm.

| **Algorithm 1:** Proposed SIF-based iterative algorithm. |
|---|
| 1: **Initialized the transmission power and Lagrangian multipliers** $\mathbf{p}^{(0)},\lambda_1^{(0)},\lambda_2^{(0)}$. |
| 2: **while** $\rho_{m,z}^{(n+1)} - \rho_{m,z}^{(n)} < \tau$ **do** |
| 3: **if** ($d_z < \frac{D}{2}$): |
| 4:   **for** $z=1:Z$ **do** $$\rho_{m,z} = \left(\frac{1}{\lambda_1} - \frac{1}{c_z}\right)^+$$ |
| 5:   **End for** |
| 6: **End if** |

**7: if** ( $d_z > \frac{D}{2}$ )

8:   **for** $z=1:Z$ **do**

$$\rho_{m,z} = \left( \frac{1}{\lambda_2} - \frac{1}{c_z} \right)^+$$

9:   **End for**
10: **End if**
Update

$$\rho_{m,z}^{(n)} = \rho_{m,z}^{(n+1)}$$

$$\lambda_1^{(n+1)} = \max\left( 0, \lambda_1^{(n)} - \theta_1 \left( \alpha P_T - P_{\max} \sum_{z:d_z < \frac{D}{2}} \rho_{m,z} \right) \right)$$

$$\lambda_2^{(n+1)} = \max\left( 0, \lambda_2^{(n)} - \theta_2 \left( (1-\alpha)P_T - P_{\max} \sum_{z:d_z > \frac{D}{2}} \rho_{m,z} \right) \right)$$

$n = n+1$.
11: **end while**
12: **End**

## IV. SIMULATION RESULTS

In this section, the results of numerical simulations based on the use of computer software are obtained, shown and discussed. Table I shows the value of the important parameters used in the simulation process.

Furthermore, Table 2 shows the numerical results of the EE based on the increase in the number of base station antennas. The fourth column of this table shows the amount of improvement in the EE value of the system for all three scenarios and confirms that the proposed method performs better than related existing algorithm.

Figure 1 shows how the total transmission power of users behaves with increasing number of antennas. As the number of the BS antennas increases, the total transmission power of the users decreases.

As shown in the figure, the proposed method performs better than the existing algorithm.

**Table 1.** Simulation Parameters.

| Parameters | Value |
|---|---|
| RB bandwidth | 120 kHz |
| Number of transmission antennas | 128 |
| Number of users | 3 |
| Noise spectral density | -175 dBm/Hz |
| Factor $\varphi$ (related to the channel matrix) | 1 |

**Table 2**. The EE versus the number of antennas.

| Number of antennas | EE convergence propose algorithm (Mbit/J) | EE convergence existing algorithm (Mbit/J) | Amount of improvement (Mbit/J) |
|---|---|---|---|
| 150 | 3.035 | 2.125 | 0.910 |
| 220 | 2.713 | 1.894 | 0.819 |
| 345 | 1.865 | 1.462 | 0.403 |

To better understand that the proposed method performs better than the proposed method in [7], we express the difference between the two points on this figure numerically. Performance of the proposed method has been improved for the number of base station antennas equal to 200 by 0.02 mW to 0.03 mW.

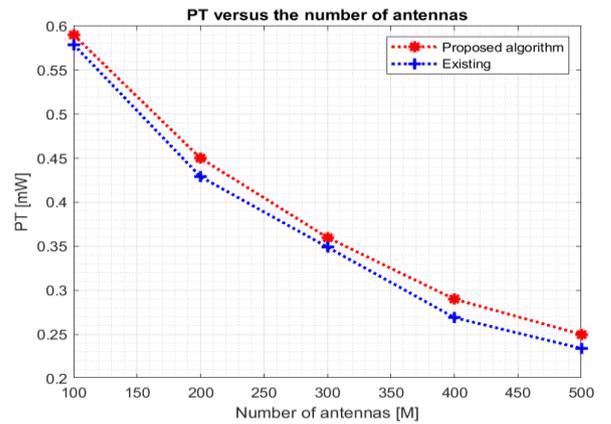

Fig. 1. $P_T$ versus number of antennas.

Figure 2 clearly shows how the EE value of the system changes based on the increase of the constant power of the circuit. As it is shown in this figure and based on the definition of the EE, by increasing constant circuit power, the EE value decreases. In addition, Figure 2 shows an improvement in the performance of the proposed scheme in comparison with [7].

For instance, at constant power 4 dBm, the proposed scheme is improved by 0.61 M bit/j and at constant power 6 dBm by 0.34 M bit/j.

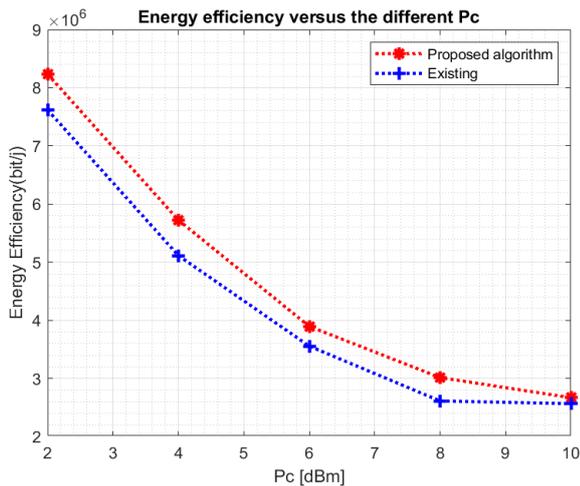

Fig. 2. The EE versus different values of $P_c$.

## V. CONCLUSION

In this article, a new efficient algorithm is proposed for the power allocation in the massive MIMO systems based on the water-filling strategy and cell division technique. In our proposed scheme, by exploiting water-filling scheme in a cell with two zones, we allocate more power to the farther users and to the users which have good channel condition. We obtained an iterative and low-complexity scheme which solves the PA problem efficiently. Based on the simulation and numerical results of the implementation of this PA scheme, we evaluate the performance of the proposed scheme. The results clearly support our claim that the proposed scheme outperforms other existing algorithms.

REFERENCES

[1] D. Gesbert, M. Kountouris, R. W. Heath, C.-B. Chae, and T. Salzer, "Shifting the MIMO paradigm," IEEE signal processing magazine, vol. 24, pp. 36-46, 2007.

[2] E. G. Larsson, O. Edfors, F. Tufvesson, and T. L. Marzetta, "Massive MIMO for next generation wireless systems," *IEEE communications magazine,* vol. 52, pp. 186-195, 2014.

[3] S. H. Mousavi and J. Pourrostam, "Fast Updating the STBC Decoder Matrices in the Uplink of a Massive MIMO System.", ICCE 2018.

[4] A. Sakhaei Gharagezlou, M. Nangir, N. Imani, and E. Mirhosseini, "Energy Efficient Power Allocation in Massive MIMO Systems with Power Limited Users. ", In 2020 4th International Conference on Telecommunications and Communication Engineering (ICTCE), Springer 2020.

[5] T. Manglayev, R.C. Kizilirmak, and Y.H. Kho, "Optimum power allocation for non-orthogonal multiple access (NOMA)." 2016 IEEE 10th International Conference on Application of Information and Communication Technologies (AICT). IEEE, 2016.

[6] J. Zhang, Y. Jiang, P. Li, F. Zheng, and X. You, "Energy efficient power allocation in massive MIMO systems based on standard interference function." 2016 IEEE 83rd Vehicular Technology Conference (VTC Spring). IEEE, 2016.

[7] A. Sakhaei Gharagezlou, J. Pourrostam, M. Nangir, and M. M. Safari. "Energy Efficient Power Allocation in Massive MIMO NOMA Systems Based on SIF Using Cell Division Technique." In 2020 2nd Global Power, Energy and Communication Conference (GPECOM), pp. 356-361. IEEE, 2020.

[8] M. Zeng, A. Yadav, O. A. Dobre, and H. V. Poor. "Energy-efficient power allocation for MIMO-NOMA with multiple users in a cluster." IEEE Access 6 (2018): 5170-5181

[9] D. P. Palomar and J. R. Fonollosa, "Practical algorithms for a family of waterfilling solutions," IEEE transactions on Signal Processing, vol. 53, pp. 686-695, 2005.

[10] S. Yu, G. Daoxing, L. Lu, and D. Xiaopei, "A modified water-filling algorithm of power allocation," in 2016 IEEE Information Technology, Networking, Electronic and Automation Control Conference, 2016, pp. 1125-1129.

[11] A. Jadhav, S. Mujawar, and P. Pise, "Optimal and water-filling algorithm approach for power allocation in OFDM based cognitive radio system," International Journal of Engineering Research and Technology, vol. 10, 2017.

[12] Q. Qi, A. Minturn, and Y. Yang, "An efficient water-filling algorithm for power allocation in OFDM-based cognitive radio systems," in *2012 International Conference on Systems and Informatics (ICSAI2012)*, 2012, pp. 2069-2073.

[13] M. N. Shahida, R. Nordin, and M. Ismail, "An improved water-filling algorithm based on power allocation in network MIMO," Telecommunication Systems, vol. 75, pp. 447-460, 2020.

[14] Z. Ding, F. Adachi, and H. V. Poor, ''The application of MIMO to nonorthogonal multiple access,'' IEEE Trans. Wireless Commun., vol. 15, no. 1, pp. 537–552, Jan. 2016.